\documentclass[11pt,preprint]{aastex}
\usepackage{epsfig,color}
\usepackage{mathrsfs}
\usepackage{ulem}
\usepackage{epstopdf}
\usepackage{hyperref}
\hypersetup{colorlinks=True,citecolor=blue}

\DeclareMathAlphabet{\mathbi}{OT1}{ptm}{bx}{it}
\SetMathAlphabet\mathbi{bold}{OT1}{ptm}{bx}{it}

\def\mathdotM{\dot{\mathscr{M}}}
\def\mbh{M_{\bullet}}
\def\rblr{R_{_{\rm BLR}}}

\def\sunm{M_{\odot}}
\def\civ{C~{\sc iv}}
\def\oiii{\rm [O~{\sc iii}]}
\def\oiiid{\rm [O~{\sc iii}]~$\lambda\lambda4959,5007$}
\def\oiiis{\rm [O~{\sc iii}]~$\lambda5007$}
\def\ergs{{\rm erg~s^{-1}}}
\def\feii{\rm Fe~{\sc ii}}
\def\fblr{f_{_{\rm BLR}}}
\def\heii{\rm He~{\sc ii}~$\lambda4686$}
\def\kms{km~s$^{-1}$}
\def\ergs{\rm erg~s^{-1}}
\def\VP{{\rm VP}}

\begin{document}

\title{Reverberation Mapping of the Broad-line Region in NGC~5548: Evidence for Radiation Pressure?}

\author
{Kai-Xing Lu\altaffilmark{1,2},
Pu Du\altaffilmark{2},
Chen Hu\altaffilmark{2},  
Yan-Rong Li\altaffilmark{2}, 
Zhi-Xiang Zhang\altaffilmark{2}, 
Kai Wang\altaffilmark{2},
Ying-Ke Huang\altaffilmark{2},
Shao-Lan Bi\altaffilmark{1}, 
Jin-Ming Bai\altaffilmark{3},
Luis C. Ho\altaffilmark{4,5} and
Jian-Min Wang\altaffilmark{2,6,*} 
}

\altaffiltext{1}
{Department of Astronomy, Beijing Normal University, Beijing 100875, China}

\altaffiltext{2}
{Key Laboratory for Particle Astrophysics, Institute of High Energy Physics,
Chinese Academy of Sciences, 19B Yuquan Road, Beijing 100049, China.}

\altaffiltext{3}
{Yunnan Observatory, Chinese Academy of Sciences, Kunming 650011, Yunnan, China.}
 
\altaffiltext{4}
{Kavli Institute for Astronomy and Astrophysics, Peking University, Beijing 100871, China} 

\altaffiltext{5}
{Department of Astronomy, School of Physics, Peking University, Beijing 100871, China} 

\altaffiltext{6}
{National Astronomical Observatories of China, Chinese Academy of Sciences,
 20A Datun Road, Beijing 100020, China}

\altaffiltext{*}
{Corresponding author: wangjm@ihep.ac.cn}

\begin{abstract}
NGC~5548 is the best-observed reverberation-mapped active galactic nucleus 
with long-term, intensive monitoring. Here we report results from a new observational campaign between January and July, 2015.
We measure the centroid time lag of the broad H$\beta$ emission line with respect to the 5100 \AA\, continuum 
and obtain $\tau_{\rm cent} = 7.20^{+1.33}_{-0.35}$ days in the rest frame.
This yields a black hole mass of $\mbh=8.71^{+3.21}_{-2.61} \times 10^{7}M_{\odot}$
using a broad H$\beta$ line dispersion of $3124\pm302$ \kms\ and a 
virial factor of $\fblr=6.3\pm1.5$ for the broad-line region (BLR), 
consistent with the mass measurements from previous H$\beta$ campaigns. 
The high-quality data allow us to construct a velocity-binned delay map for
the broad H$\beta$ line, which shows a symmetric response pattern around the line center,
a plausible kinematic signature of virialized motion of the BLR. 
Combining all the available measurements of H$\beta$ time lags and 
the associated mean 5100~{\AA} luminosities over 18 campaigns between 1989 and 2015, 
we find that the H$\beta$ BLR size varies with the mean optical luminosity, 
but, interestingly, with a possible delay of $2.35_{-1.25}^{+3.47}$\,yrs.
This delay coincides with the typical BLR dynamical timescale of NGC~5548, 
indicating that the BLR undergoes dynamical changes, possibly driven by radiation pressure. 
\end{abstract}


\section{Introduction}
Reverberation mapping (RM) is a powerful tool to probe the geometry and structure
of broad-line regions (BLRs) in active galactic nuclei (AGNs) (\citealt{Bahcall1972, Blandford1982, Peterson1993}). 
Over the past four decades,
great efforts on RM monitoring have yielded a precious sample of $\sim60$ nearby Seyfert galaxies
and quasars with measurements of H$\beta$ time lags (e.g., \citealt{Bentz2013, Du2016a}), 
among them NGC~5548, the best-observed source
that has been intensively monitored by 17 individual RM campaigns,
including the recent Space Telescope and Optical Reverberation Mapping Project
(AGN STORM; \citealt{DeRosa2015, Edelson2015, Fausnaugh2015arXiv}; see \citealt{Peterson2002} 
for a summary of the first 13 campaigns; \citealt{Bentz2007, Bentz2009}). 
NGC~5548 therefore serves as a valuable laboratory to study in detail the long-term 
variations of the BLR (\citealt{Wanders1996, Sergeev2007}), as well as the consistency and reliability 
of RM-based black hole (BH) mass measurements (\citealt{Peterson1999, Collin2006}).

NGC 5548 follows the relation $\rblr\propto L_{5100}^{0.79\pm0.2}$ 
(\citealt{Eser2015}), where $L_{5100}$ is the optical luminosity at 5100 \AA. This relation for 
NGC 5548 is significantly different from the well-known radius$-$luminosity relation
$\rblr\propto L_{5100}^{0.53_{-0.03}^{+0.04}}$ for the overall RM sample
(\citealt{Kaspi2000}; \citealt{Bentz2013}). This difference needs to be understood. 
On the other hand, the geometry and kinematics of the BLR in NGC 5548 have been investigated by 
velocity-resolved mapping in several studies (e.g., 
\citealt{Denney2009b}; \citealt{Bentz2010}; \citealt{DeRosa2015}), and by recently developed
dynamical modelling \citep{Pancoast2014} using the data taken by the 2008 Lick AGN Monitoring Project 
(LAMP; \citealt{Bentz2009}).  However, the inferred BLR dynamics seems diverse, and there is 
no consensus\footnote{This can be seen by comparing Figure 3 from Denney 
et al. (2009b) with Figure 19 from Bentz et al. (2010), who present observations from 2007 and 2008, respectively.
Such a difference cannot be caused by intrinsic variations of 
the BLR because the time separation between the two campaigns is significantly shorter than the BLR
dynamical timescale (see Equation 9).}. 

To investigate the above issues, we conducted a new observational campaign for NGC 5548 in 2015. 
This paper presents the results of our new RM campaign. 
In Section 2, we describe the observations and the data reduction in detail.
In Section 3, we perform the time series analysis and measure the H$\beta$ time lags and
construct the velocity-resolved lags of the broad $\rm H\beta$ line. 
We investigate the structure and dynamics of the BLR in Section 4, and discuss the BH mass 
measurements, accretion rates, and the long-term variations of BLR size in Section 5. We draw our
conclusions in Section 6. Throughout the paper, a cosmology with $H_0=67{\rm
~km~s^{-1}~Mpc^{-1}}$, $\Omega_{\Lambda}=0.68$, and $\Omega_{\rm M}=0.32$ is
adopted (\citealt{Ade2014}). 

\section{Observations and Data Reduction}

\subsection{Observations}
The spectroscopic and photometric observations of NGC~5548 were made using the Yunnan Faint Object 
Spectrograph and Camera (YFOSC), mounted on the Lijiang 2.4 m telescope at Yunnan Observatory,
Chinese Academy of Sciences.
Working at the Cassegrain focus, YFOSC is a versatile instrument for low-resolution spectroscopy and photometry. 
It is equipped with a  back-illuminated 2048$\times$2048 pixel CCD, with pixel size 13.5 $\mu$m, 
pixel scale 0.283$''$ per pixel, and field-of-view $10'\times10'$. YFOSC can automatically switch 
from spectroscopy to photometry within 1 s (see \citealt{Du2014}).

\begin{figure}[t!]
\begin{center}
\includegraphics[angle=0,width=0.45\textwidth]{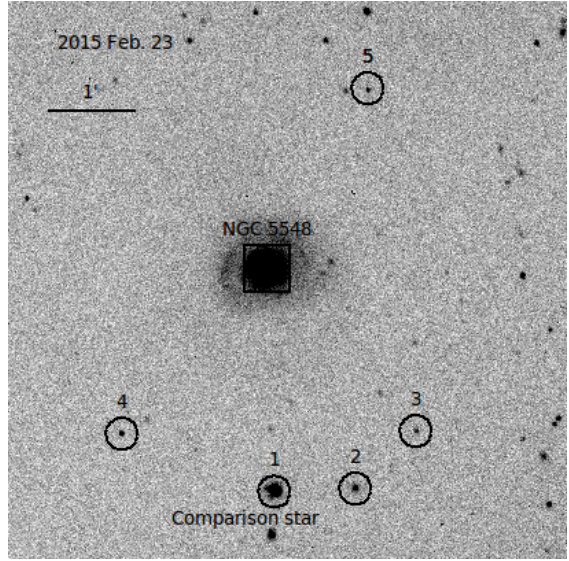}
\end{center}
\caption{\footnotesize
Johnson $V$-band image of NGC 5548 from the Lijiang 2.4m telescope, observed 
on February 23, 2015. NGC 5548 is located in the center of the field. 
Stars 1$-$5 are selected as photometric comparison stars, and star 1 is selected as the spectral comparison star.
}
\label{fig:img}
\end{figure}

Our observations started on January 7, 2015 and ended on July 11, 2015. 
To get an accurate flux calibration, 
we simultaneously observed a nearby comparison star along the long-slit as a reference standard.
Such an observation strategy was described in detail by \citet{Maoz1990} and \citet{Kaspi2000}, 
and was recently adopted by \citet{Du2014}.
As shown in Figure~\ref{fig:img}, we chose the star labeled ``No.1'' as the comparison star. 
Given the seeing of $1.0^{\prime\prime}-2.5^{\prime\prime}$  throughout the monitoring period, 
we fixed the projected slit width at $2.5''$.  We
used Grism 14, which provides a resolution of 92 \AA\, mm$^{-1}$ (1.8 \AA\, pixel$^{-1}$) 
and covers the wavelength range 3800$-$7200 \AA\,.
Standard neon and helium lamps were used for wavelength calibration. 
To reduce atmospheric differential refraction, 
we limited the observations to airmasses $\lesssim 1.2$. This guarantees an
atmospheric refraction $\le 0.3^{\prime\prime}$ over the wavelength range $4500-5500$ \AA\, (\citealt{Filippenko1982}). 
The mean airmass for all the spectra is $1.07$, so that
any offset of the target from the slit center due to atmospheric refraction is limited to 
$\lesssim 0.14^{\prime\prime}-0.21^{\prime\prime}$, which has a negligible impact on our analysis. 

To verify the calibration of the spectroscopic data, we also made photometric observations
using a Johnson $V$ filter.  We took three consecutive exposures of 90 s each.
In total, we obtained 62 spectroscopic observations and 61 photometric observations, spanning a 
time period of 180 days. The typical cadence is $\sim 3.4$ days. 

\subsection{Data Reduction}
The two-dimensional spectroscopic images were reduced 
using the standard {\tt IRAF} tools before absolute flux calibration.
This includes bias subtraction, flat-field correction, and wavelength calibration. 
All the spectra were extracted using a uniform aperture of 30 pixels (8.5$^{\prime\prime}$), and
the background was determined from two adjacent regions on either side of the aperture region. 
As described in \citet{Du2014}, absolute flux calibration was done in two steps. 
(1) The observations taken during nights with good weather conditions were used to 
calibrate the absolute flux of the comparison star, 
which was then used as the fiducial spectrum for absolute flux standard for the science observations.
(2) For each object/comparison star pair, 
a wavelength-dependent sensitivity function was obtained by comparing the star's spectrum 
to the fiducial spectrum. Then this sensitivity function was applied to  
calibrate the observed spectrum of the target. 
The spectra calibrated in this way show a small fluctuation of the \oiiis\ flux at a level of 2\%,
which can be regarded as the accuracy of our absolute flux calibration. 

The $V$-band images were also reduced using standard {\tt IRAF (V2.16)} procedures. Instrumental
magnitudes were measured with respect to five selected reference stars in the field
(see Figure \ref{fig:img}). The typical accuracy of the photometry is 0.015 mag. 

\subsection{Spectral Measurements}
\label{sec:lc}

Following \citet{Hu2015}, 
we measured the 5100~{\AA} continuum and broad H$\beta$ line using a spectral
fitting scheme. The fitting components include 
(1) a single power-law continuum; 
(2) a stellar component for the host galaxy;
(3) \feii\, emission;
(4) broad H$\beta$ emission line; 
(5) broad \heii\, emission line; and 
(6) narrow emission lines of \oiiid, \heii, H$\beta$, and several coronal lines (such as 
[Fe {\sc vii}] $\lambda5158$, [Fe {\sc vi}] $\lambda5176$ and [Ca {\sc v}] $\lambda5309$; see the details 
in \citealt{Hu2015}).
The spectral template for the host galaxy is a stellar population model with an age of 11 Gyr 
and a metallicity $Z=0.05$ \citealt{Bruzual2003}. 

\begin{figure}[t!]
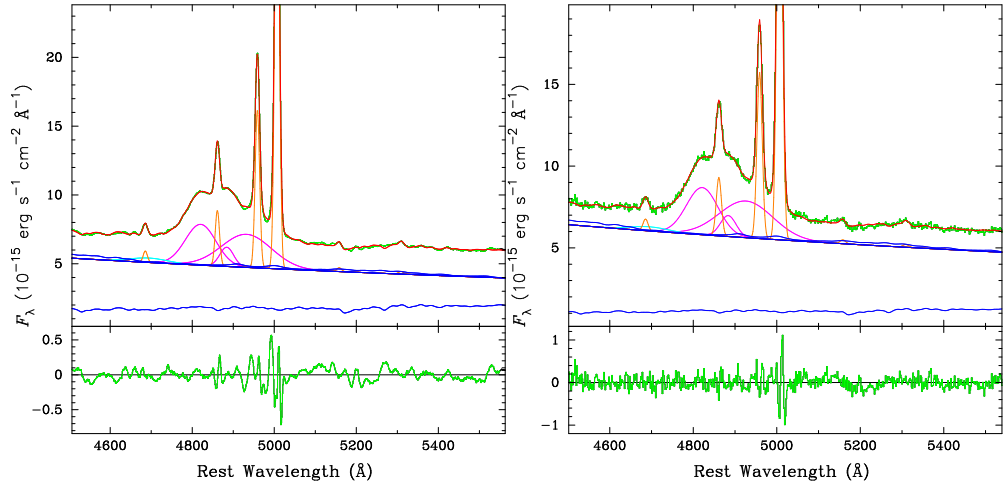

\begin{center}
{\includegraphics[angle=-90,width=0.4\textwidth]{fig2a.eps}}
{\includegraphics[angle=-90,width=0.4\textwidth]{fig2b.eps}}
\end{center}
\caption{\footnotesize
Multi-components fitting of ({\it left}) the mean spectrum and ({\it right})
an individual spectrum of NGC~5548. 
The trace shows the spectrum corrected for Galactic extinction (green) 
and the best-fit model (red), which is composed of the AGN power-law continuum (blue), 
\feii\, emission lines (blue; template from \citealt{Boroson1992}), host galaxy (blue), broad H$\beta$ (magenta), 
broad \heii\, (cyan), and several narrow emission lines (orange). 
The bottom trace shows the residuals. 
}
\label{fig:spec_fit}
\end{figure}

During the observations, seeing variations and mis-centering of the object in the slit 
lead to varying amounts of host galaxy light in the final spectrum.  To remove this effect, the
flux of the host galaxy component is set to a free parameter in our fitting scheme. 
During the whole monitoring period, the $\rm H\beta$ emission line of NGC~5548 showed extreme broad wings, 
in addition to a strong narrow component. We slightly modified the fitting scheme of \citet{Hu2015}
in two aspects, by 
adding a narrow component to $\rm H\beta$ and 
using three Gaussians to fit the broad component.
Figure~\ref{fig:spec_fit} shows the fitting results for the mean spectrum
of our campaign and for an individual spectrum.

We measured the 5100~\AA\, continuum from the best-fit power-law component and 
the $\rm H\beta$ flux by integrating the best-fit broad components 
of $\rm H\beta$ from $4710$\,\AA\, to $5050$\,\AA.  
For the purpose of constructing the velocity-resolved delay map (see Section~\ref{sec:vrm}), 
we obtained the broad $\rm H\beta$ profile by subtracting the host galaxy, \feii\, and \oiii\, emission lines, 
and the narrow component of H$\beta$.  Table~\ref{tab:lc_dat} summarizes the light curves. 
Figure~\ref{fig:lc_ccf} plots the light curves of the $V-$band photometry 
(instrumental magnitude in an arbitrary unit), 5100\,\AA\, continuum, and broad $\rm H\beta$ flux. 

\section{Time Series Analysis}
\subsection{Variability Characteristics}
Following standard practice (e.g., \citealt{Rodriguez-Pascual1997}), we calculate the variability amplitude
of the light curves of the 5100~{\AA} continuum 
and H$\beta$ emission line by 
\begin{equation}
F_{\rm var}=\frac{\left(\sigma^2-\Delta^2\right)^{1/2}}{\langle F\rangle} \, 
\end{equation}
and its uncertainty (\citealt{Edelson2002})
\begin{equation}
\sigma_{_{F_{\rm var}}} = \frac {1} {F_{\rm var}} \left(\frac {1}{2 N}\right)^{1/2} \frac {\sigma^2}{\langle F\rangle} \, ,
\end{equation}
where $\langle F\rangle=N^{-1}\sum_{i=1}^NF_i$ is the average flux, $F_i$ is the flux 
of the $i$-th observation of the light curve, $N$ is the total number of observations,
$\sigma^2=\sum_{i=1}^N\left(F_i-\langle F\rangle\right)^2/(N-1)$, 
$\Delta^2=\sum_{i=1}^N\Delta_i^2/N$,
and $\Delta_i$ is the uncertainty of $F_i$. Table~\ref{tab:lcst}  lists the statistics 
of the light curves. The variability amplitudes of the 5100~{\AA} continuum and broad H$\beta$ line
are $F_{\rm var}=0.23$ and 0.10, respectively, which are generally comparable with those of previous RM campaigns,
after correcting for host galaxy and narrow H$\beta$ contributions (\citealt{Peterson2002, Bentz2007, Denney2010}). 
We also calculate another standard measure of variability, $R_{\rm max}$, simply
defined by the ratio of maximum to minimum flux.  The values of $R_{\rm max}$
are 2.31 and 1.52 for the 5100~{\AA} continuum and broad H$\beta$ line, respectively.

\subsection{H$\beta$ Time Lags} 
\label{tdr}
We compute the time lag of H$\beta$ line flux relative to 5100~{\AA} continuum 
using the interpolation cross-correlation function method 
(ICCF; \citealt{Gaskell1986}; \citealt{Gaskell1987}; \citealt{White1994}). 
The time lag is measured by two standard approaches: the peak location $\tau_{\rm peak}$ of the ICCF
($r_{\rm max}$) and the centroid $\tau_{\rm cent}$ of the ICCF around the peak above a typical value ($r\ge0.8~r_{\rm max}$). 
Their respective uncertainties  were obtained using the Monte Carlo 
``flux randomization$/$random subset sampling'' method 
described by \citet{Peterson1998} and \citet{Peterson2004}. 
The Monte Carlo simulations were run with 1000 realizations, and  
the distributions of the peak and centroid (CCPD and CCCD) were created from the generated samples. 
The uncertainties of $\tau_{\rm peak}$ and $\tau_{\rm cent}$ 
are then calculated from the CCPD and CCCD, respectively, with a 68.3\% confidence level ($1\sigma$). 

\begin{figure}[t!]
\begin{center}
\includegraphics[angle=0,width=0.8\textwidth]{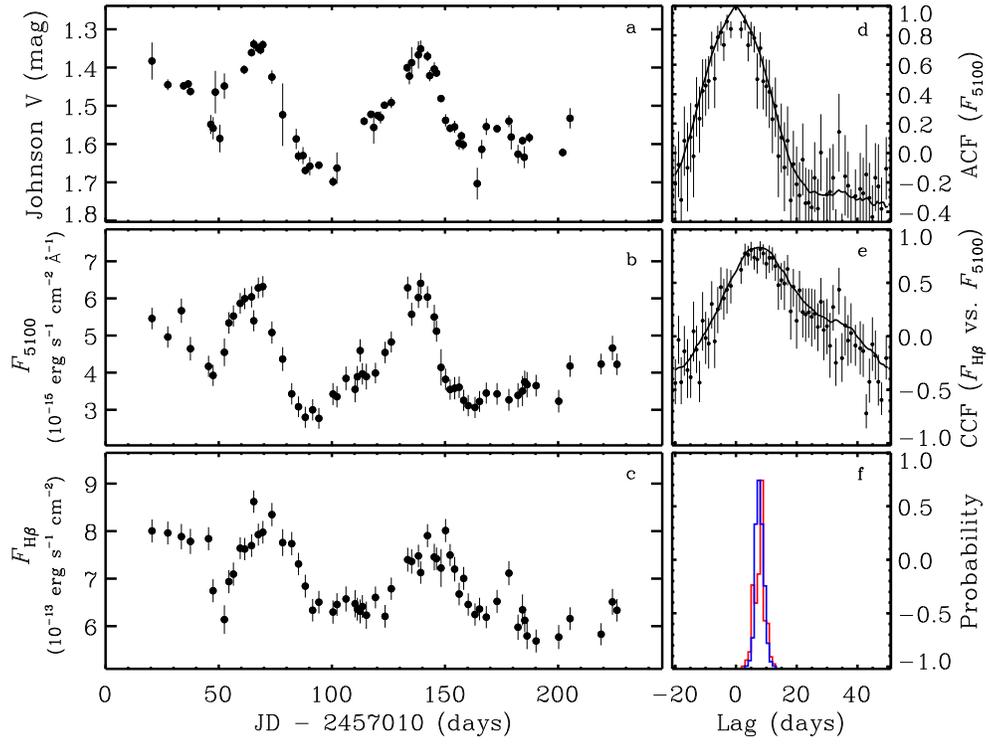}
\end{center}
\caption{\footnotesize
Light curves and  the results of cross correlation analysis.  
Panels ({\it d, e, f}) are the ACF of continuum at 5100 \AA\,, 
CCF between the $\rm H\beta$ emission line and continuum at 5100 \AA\,, 
and the Monte Carlo simulations of peak (red) and centroid (blue) of lags, respectively. 
In panels ({\it d, e}), the solid lines show the ICCF, and 
points with error bars show the ZDCF. 
The light curves in panels ({\it b, c}) include the systematic uncertainties. 
}
\label{fig:lc_ccf}
\end{figure}

In Figures~\ref{fig:lc_ccf}($d$) and \ref{fig:lc_ccf}($e$), we show the auto cross-correlation function (ACF) 
of the light curve of 5100~{\AA} continuum, the ICCF between the H$\beta$ flux and 
5100~{\AA} continuum, and the CCPD and CCCD of the ICCF, respectively.
We find that the ICCF peak occurs at $\tau_{\rm peak}=7.20^{+1.33}_{-0.35}$ days ($r_{\rm max}=0.83$) and the 
ICCF centroid occurs at $\tau_{\rm cent}=7.18^{+1.38}_{-0.70}$ days in the rest frame (see Table~\ref{tab:rm_p}). 
As an independent check, we also calculate the Z-transformed discrete correlation function 
(ZDCF; \citealt{Edelson1988}; \citealt{Alexander1997}) and  superpose the corresponding ZDCF in 
Figures \ref{fig:lc_ccf}($d$) and \ref{fig:lc_ccf}($e$).  
As can be seen, the ZDCFs are in good agreement with the ICCFs.

\subsection{Velocity-resolved Reverberation Mapping}
\label{sec:vrm}
\begin{figure}[t!]
\begin{center}
\includegraphics[angle=0,width=0.65\textwidth]{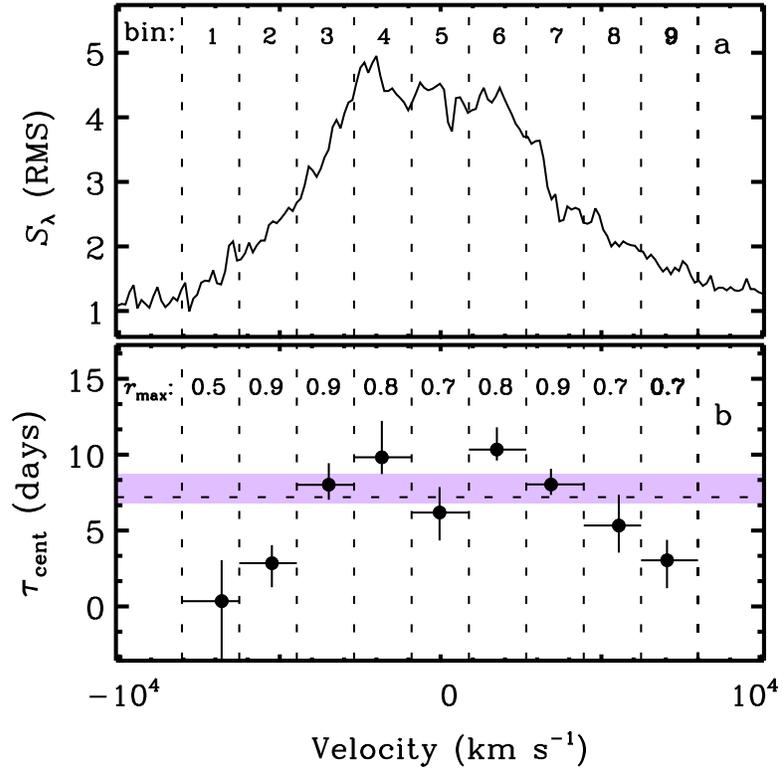}
\end{center}
\caption{\footnotesize
Panel ({\it a}) shows the RMS spectra of H$\beta$. We divided the profile into 9 bins.
Panel ({\it b}) shows the centroid lags of each H$\beta$ bin.
The vertical dash-lines are the edges of bins, and the black horizontal dashed line with purple 
shaded area shows the average time lag with uncertainties.  
}
\label{fig:2d_tv}
\end{figure}

Velocity-resolved RM is widely used to reveal the kinematic signatures of 
BLRs (e.g., \citealt{Grier2013}; \citealt{Bentz2008}, \citealt{Bentz2010};
\citealt{Denney2009a,Denney2009b,Denney2010}; \citealt{DeRosa2015}; \citealt{Du2016b}). 
The high-quality spectroscopic data of our campaign allow us to construct the velocity-binned 
delay map of the H$\beta$ line. Our procedure is as follows. 
Using the method described in the Appendix, we first calculate the RMS spectrum of the broad 
H$\beta$ profiles obtained in Section \ref{sec:lc}. As illustrated in Figure~\ref{fig:2d_tv}({\it a}),
we then select a wavelength range from 4731 \AA to 4991 \AA\ in the rest frame 
and divide the $\rm H\beta$ profiles into nine uniformly spaced bins (each bin has a  
velocity width of $\sim 1700\,{\rm  km~s^{-1}}$)\footnote{It should be pointed out
that the instrumental broadening
is about 500~${\rm km~s^{-1}}$ for the $2.5^{\prime\prime}$ slit, significantly smaller
than the width of the velocity bin. We thus did not employ the method of Du et al. (2016b) to correct
the spectra for instrumental broadening.}. 
The light curve of each bin is finally obtained by just integrating the flux in the bin. 
The time lag of each bin and the associated uncertainties are determined using the 
same procedures as described in Section~\ref{tdr}. 

\begin{figure}[t!]
\begin{center}
\includegraphics[angle=0,width=0.7\textwidth]{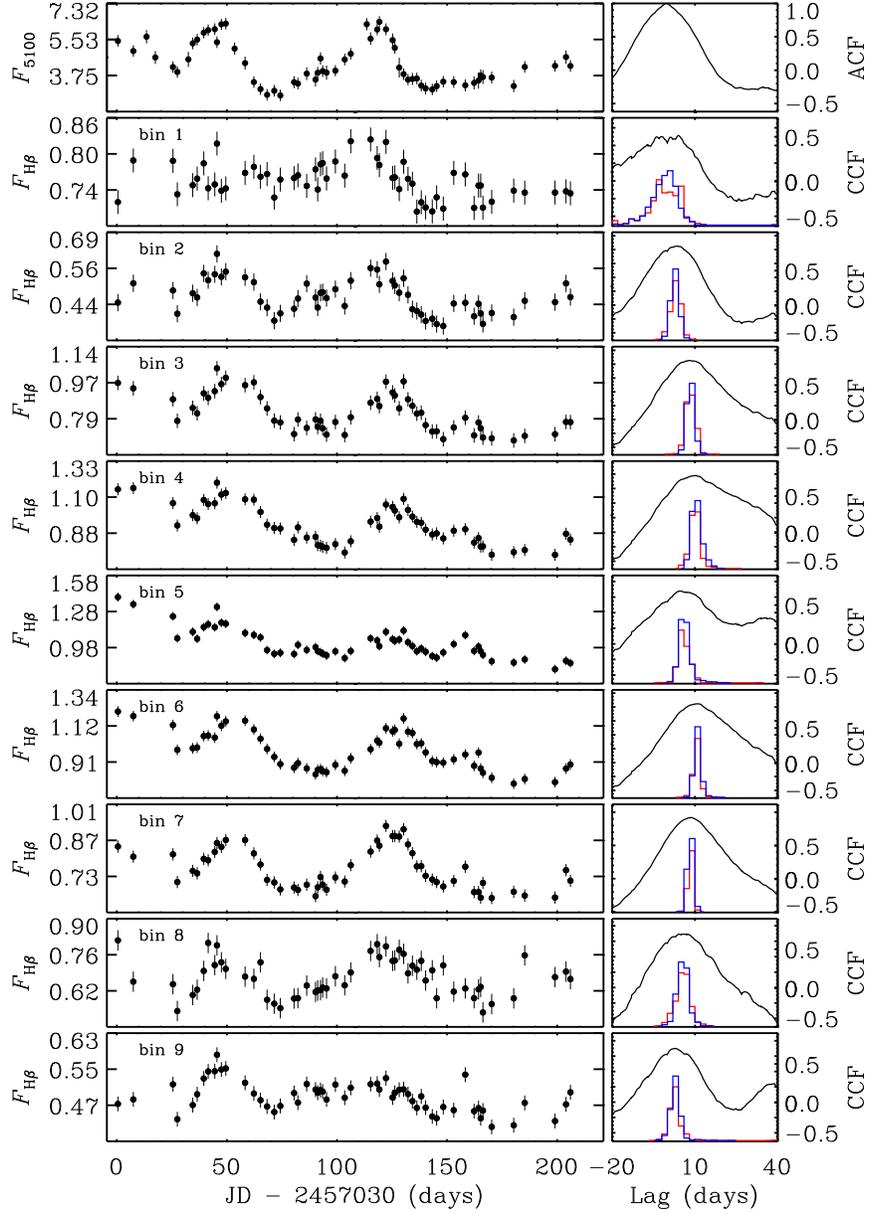}
\end{center}
\caption{\footnotesize
Velocity-resolved reverberation mapping. 
The left panels show the light curves of continuum at 5100 \AA, 
and $\rm H\beta$ emission line of each velocity bin, respectively. 
We divided the H$\beta$ profile into 9 velocity bins and numbered each bin from 1 to 9. 
The right panels correspond to the ACF of continuum and the
CCF between the light curve of each bin and continuum, respectively.
Monte Carlo simulations of the peaks (red) and centroid (blue) 
of time lags are overplotted in the CCF panels.
}
\label{fig:2d_lc}
\end{figure}

In Figure~\ref{fig:2d_lc}, we plot the obtained light curves of the nine bins, along with the light curve of the
5100~{\AA} continuum for the sake of comparison. 
The corresponding ICCFs between the light curve of each bin and the continuum are shown in the right panels
of Figure~\ref{fig:2d_lc}, in which the CCCD and CCPD are also superposed. 
The velocity-resolved delay map is plotted in Figure~\ref{fig:2d_tv}({\it b}). We can find that  
the delay map has a symmetric pattern, with longer response at the line 
core and shorter response at the wings (except for bin 5). 

Velocity-resolved delay maps of the broad H$\beta$ line in NGC~5548 were derived previously
in the MDM campaign undertaken in 2007 (\citealt{Denney2009a,Denney2009b}) and  
in the LAMP campaign in 2008 (\citealt{Bentz2009}).
The delay map of our campaign is very similar to that of the MDM campaign. \cite{Denney2009a} 
conclude that the symmetric delay map of NGC 5548 indicates that there is no radial gas motion in the BLR. 
The data quality of the LAMP campaign is relatively poor so that the resulting delay map in \cite{Bentz2009}
does not reveal clear signatures for the BLR motion. However, \cite{Pancoast2014} carried out 
dynamical modeling of the LAMP data and found that a narrow thick-disk-like BLR geometry with dominant inflows 
can explain the variations of the broad H$\beta$ line. 
On the other hand, it is worth mentioning that 
the \civ\, velocity-resolved delay map of the AGN STORM campaign derived by
\cite{DeRosa2015} shows a symmetric structure. 
In the future, detailed analysis of the present data through dynamical modeling
(e.g., \citealt{Pancoast2011,Li2013}) will better constrain the kinematics of the H$\beta$ BLR.

\section{Structure and Dynamics of the BLR}
This section examines the virial assumption for the BLR motions and the relation
between the H$\beta$ BLR size and optical luminosity, by combining data from all the available 
RM campaigns of NGC~5548. We compile the width and time lags of $\rm H\beta$ 
from the literature (\citealt{Peterson2004}; \citealt{Collin2006}; \citealt{Bentz2007}; 
\citealt{Denney2010}; \citealt{Bentz2010}; \citealt{Fausnaugh2015arXiv}). 
\citet{Eser2015} re-calibrated  the 5100 \AA\, flux using the updated flux measurements
of the host galaxy given in \citet{Bentz2013}. We directly take these re-calibrated fluxes to calculate 
the 5100~{\AA} luminosity. Table~\ref{tab:tot} summarizes
all the RM measurements of H$\beta$ lag, H$\beta$ flux, and
the dispersion and full width half maximum (FWHM) velocity of the H$\beta$ line 
from RMS and mean spectra.

\subsection{The Virial Relation}
\label{sec:vp}
We employ the method of \citet{Peterson2004} to measure the  FWHM and $\sigma_{\rm line}$ of $\rm H\beta$ 
from the mean and RMS spectra (Table 3; Appendix). 
We find $\rm FWHM / \sigma_{\rm line} \approx 3$, 
which is larger than 2.35 for a Gaussian profile, indicating that the BLR has less turbulent 
motion \citep{Kollatschny2013}. We calculate the virial product (Table 4)
\begin{equation}
{\rm VP}=\frac{c\tau_{_{\rm H\beta}} V^2}{G}, 
\label{eqn3}
\end{equation}
where $\tau_{_{\rm H\beta}}$ is the H$\beta$ time lag, $c$ is the speed of light, 
$G$ is the gravitational constant, and $V$ is the line width. 
If the BLR motion is dominated by the gravity of the central BH and is virialized, 
VP should be constant, and $V\propto\tau^{-1/2}_{_{\rm H\beta}}$. 

\begin{figure}[t!]
\begin{center}
\includegraphics[angle=0,width=0.75\textwidth]{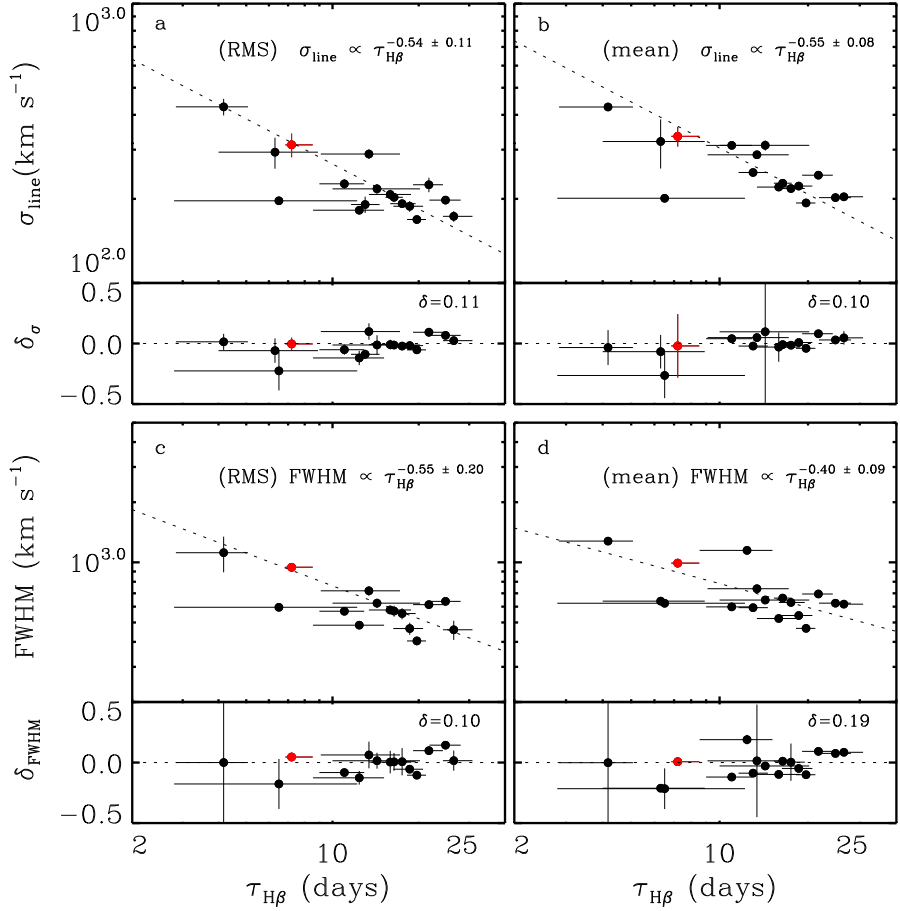}
\end{center}
\caption{\footnotesize
The dependence of $\rm H\beta$ line dispersion and FWHM on $\rm H\beta$ lag for mean and RMS spectra of NGC 5548. 
The bottom panel of each plot shows the residuals from the virial relation. 
The inset value of $\delta$ shows the deviation of the BLR motion from the virial relation, defined by Equation (4). 
%
%
The red point is from the present work.
}
\label{fig:w_t}
\end{figure}

Figure~\ref{fig:w_t} shows the relation between $V$ and $\tau_{_{\rm H\beta}}$ 
using four measures of line width, namely $\sigma_{\rm line}$ and FWHM from the mean and RMS spectra. 
The relations between $V$ and $\tau_{_{\rm H\beta}}$
have a slope of ($-0.54$, $-0.55$, $-0.55$, $-0.40$) 
for $\sigma_{\rm line}$ and FWHM of the RMS and mean spectra, respectively. 
Generally, the H$\beta$ line width obeys the virial relation within the uncertainties, 
consistent with the results reported by \citet{Peterson2004} and \citet{Bentz2007}. 
To quantitatively describe any
deviation of the BLR motion from the virial relation, we define a parameter 
\begin{equation}
\delta=\sum_i^N\delta_{V_i}^2,~~~{\rm and~~~} \delta_{V_i}=\log\left(\frac{V_i}{V_{{\rm vir}}}\right),
\end{equation}
where 
$V_{\rm vir}=\left( G\langle {\rm VP}\rangle/\rblr\right)^{1/2}$ and $\langle{\rm VP}\rangle$ is the 
average virial product. We find that the deviation $\delta=(0.11,0.10,0.10,0.19)$ for $\sigma_{\rm line}$ and FWHM of the 
RMS and mean spectra, respectively.

\subsection{The VP and 5100~{\AA} Luminosity}
\begin{figure}[t!]
\begin{center}
\includegraphics[angle=0,width=0.6\textwidth]{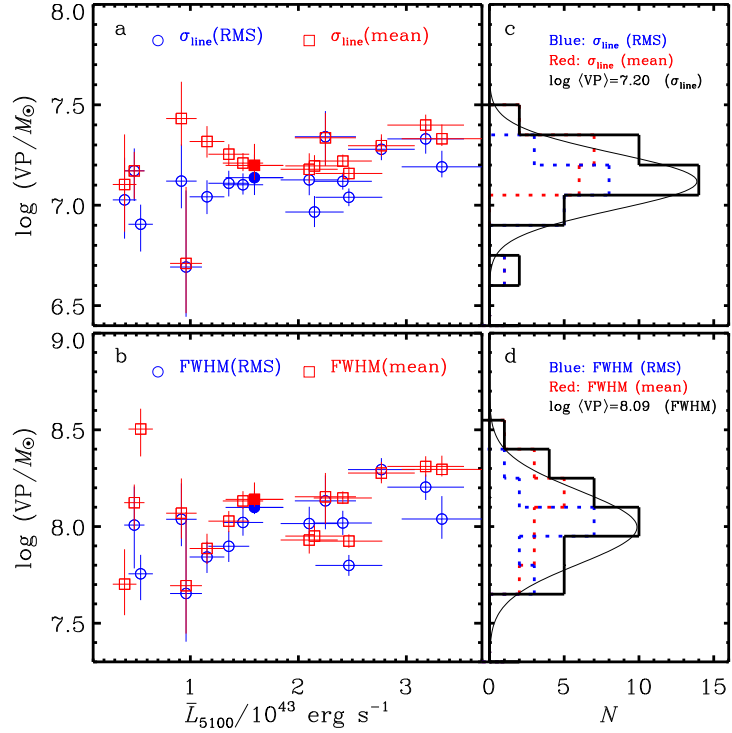}
\end{center}
\caption{\footnotesize
The virial product (VP) vs. the optical luminosity (corrected for the starlight contribution of host galaxy). 
In panels ({\it a, b}), the solid points are from this work and the open symbols are from previous RM measurements. 
Panels ({\it c, d}) show the corresponding distributions of VP, fit with a Gaussian (solid line).
}
\label{fig:f_l}
\end{figure}

Figures~\ref{fig:f_l}({\it a}) and ({\it b}) show the distribution of VP 
as a function of optical luminosity. Compared with the previous analysis of \citet{Collin2006} 
and \citet{Bentz2007}, we extend the analysis by including the latest RM measurements.
Using the procedure \texttt{FITEXY} of \cite{Press1992}, we obtain the regressions 
\begin{equation}
\log\left(\VP/\sunm\right)=\left\{\begin{array}{ll}
        (7.13\pm 0.04)+(0.21\pm0.13)\log \ell_{43}&({\rm for~ \sigma_{\rm line}}),\\ [0.8em]
        (7.97\pm 0.05)+(0.36\pm0.20)\log \ell_{43}& ({\rm for~ FWHM}),
          \end{array}\right.
\label{eqn5}
\end{equation}
where $\ell_{43}=\bar{L}_{5100}/10^{43}\,\ergs$. 
Figures~\ref{fig:f_l}({\it c}) and ({\it d}) show that the distributions 
of VP$|_{\sigma_{\rm line}}$ and VP$|_{\rm FWHM}$ have a large scatter 
(0.31 and 0.32 dex, respectively).
We note that the weak correlation between VP and luminosity is not
in conflict with the notion that the BLR is predominantly virialized. Even if 
variations in radiation pressure induces secular departures from virial equilibrium, 
the kinematics gradually adjusts to restore a quasi-virialized state. 
This test shows that the method for determining the BH
mass is robust as long as the factor $\fblr$ is reliably calibrated. 

\subsection{The BLR Radius$-$Luminosity Relation}
\begin{figure}[t!]
\begin{center}
\includegraphics[angle=0,width=0.8\textwidth]{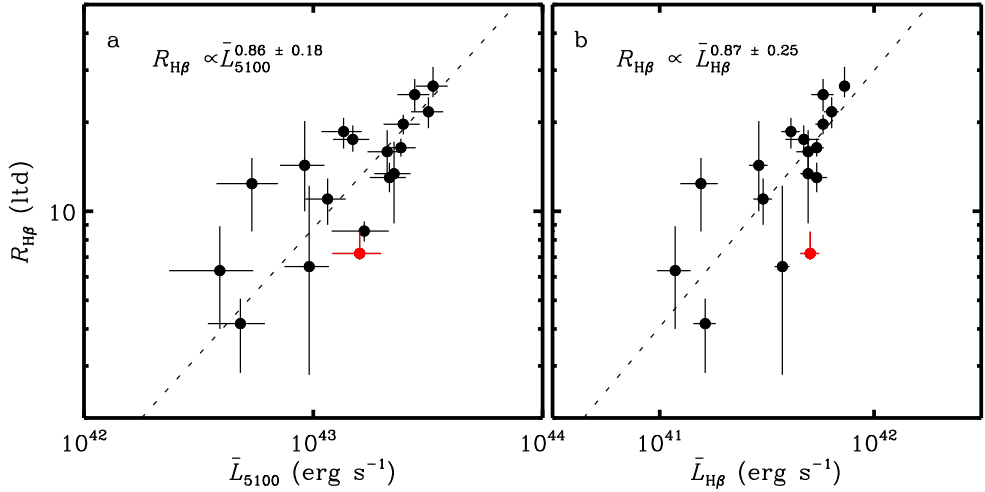}
\end{center}
\caption{\footnotesize
The ({\it a}) $R_{\rm H\beta}-\bar{L}_{\rm 5100}$ and ({\it b}) $R_{\rm H\beta}-\bar{L}_{\rm H\beta}$ relation of NGC~5548. The red point is from this work, and the black points are from previous RM measurements. 
}
\label{fig:r_l}
\end{figure}

\cite{Peterson2004} found that the $\rblr-\bar{L}_{\rm 5100}$ relation of NGC 5548 has a much steeper slope than 0.5. Figure~\ref{fig:r_l}({\it a}) reemphasises this point, by including the new measurement from our campaign.  Using \texttt{FITEXY}, we obtain  the best-fit regression of 
\begin{equation}
\log \left(R_{\rm H\beta}/{\rm ltd}\right)= (0.94\pm0.05)+(0.86\pm0.18)\log\ell_{43},
\end{equation}
where $\ell_{43}=\bar{L}_{5100}/10^{43}\,\ergs$. 
With the addition of the new data, 
our derived relation is slightly steeper than that reported 
by Kilerci Eser et al. (2015; $\rblr\propto L_{5100}^{0.79\pm0.2}$). 
Both slopes are steeper than the value of 0.5 expected from simple photoionization theory.  
For completeness, Figure~\ref{fig:r_l}({\it b}) also shows the best-fit regression 
of the relation between the H$\beta$ BLR size and the H$\beta$ luminosity, 
$R_{\rm H\beta}\propto\bar{L}_{\rm H\beta}^{0.87\pm0.25}$. 
Recently, based on data from simultaneous optical and UV observations, 
\citet{Eser2015} established the connection between the 5100~{\AA} and the 1350~{\AA} luminosity, 
as $\bar{L}_{\rm 5100}\propto\bar{L}_{\rm 1350}^{0.63\pm0.12}$.  
Using this relation, we deduce $R_{\rm H\beta} \propto \bar{L}_{\rm 1350}^{0.54\pm0.22}$, 
which is consistent with the expected slope of 0.5. 
This non-linear relation between the optical and UV emissions implies 
a complicated geometry for the accretion disk and the importance 
of radiative reprocessing (e.g., \citealt{Fausnaugh2015arXiv}; \citealt{Edelson2015}).

\section{Discussion} 
\subsection{Black Hole Mass and Accretion Rate}
Following standard practice,  we estimate the BH mass as
\begin{equation}
 \mbh = \fblr\frac{c\tau_{_{\rm H\beta}} V^{2}}{G},
\label{eqn:mass}
\end{equation}
where $\fblr$ is a coefficient that crudely accounts for the unknown 
inclination, geometry, and kinematics of the BLR. \citet{Ho2014} point out 
that the BLR in pseudobulges notably has a lower $f_{\rm BLR}$ than in classical bulges. 
For line dispersion $\sigma_{\rm line}$ measured from RMS spectra, 
$\fblr=3.2\pm0.7$ for pseudobulges, whereas $\fblr=6.3\pm1.5$ for classical bulges. 
The latter is roughly consistent, within uncertainties, 
with previous calibrations that do not take the bulge type into consideration (e.g., $\fblr=5.5\pm1.7$; \citealt{Onken2004}). 
NGC~5548 hosts a classical bulge (\citealt{Ho2014}).
Using $\fblr=6.3\pm 1.5$, we derive a BH mass of $\mbh|_{\sigma_{\rm line}}=8.71^{+3.21}_{-2.61} \times 10^7M_{\odot}$; 
line dispersion measured from mean spectra, $\fblr = 5.6\pm 1.3$ \citep{Ho2014} and 
$\mbh|_{\sigma_{\rm line}}=8.91^{+3.08}_{-2.67} \times 10^{7}M_{\odot}$. 
The two mass measurements are in good agreement. Meanwhile, 
our measurements are also consistent, within the uncertainties, 
with the overall values from previous RM campaigns. A notable exception is the work of \cite{Pancoast2014}; 
applying the BLR dynamical modeling developed by \cite{Pancoast2014a} to the LAMP data on NGC~5548, 
they obtain, without invoking the virial factor, $\mbh=3.89_{-1.49}^{+2.87}\times 10^7\sunm$, 
which is only about half of the virial-based value. 

The classical bulge of NGC~5548 has a central stellar velocity dispersion of 
$\sigma_{*}=195\pm13$ \kms\ (\citealt{Woo2010}), 
resulting in $\mbh|_{\sigma_*}=(2.75\pm 0.88)\times 10^8\sunm$ 
from the latest $\mbh-\sigma_*$ relation (\citealt{Kormendy2013}).
This is marginally larger than the virial-based BH mass estimate, after taking into account the intrinsic scatter of 
the virial factor (\citealt{Ho2014}), but significantly exceeds the BH mass determination based on dynamical modeling of the BLR
by \cite{Pancoast2014}.

With the BH mass in hand, we can calculate the dimensionless accretion rate, defined as
\begin{equation}
\mathdotM=\frac{\dot{M}_{\bullet}c^2}{L_{\rm Edd}}
         \approx 0.1\,\eta_{0.1}^{-1}\left(\frac{L_{\rm bol}}{10^{44}\,{\rm \ergs}}\right)
                                   \left(\frac{\mbh}{10^8\,\sunm}\right)^{-1}, 
\label{eqn7}
\end{equation}
where $\dot{M}_{\bullet}$ is the mass accretion rate, 
$L_{\rm Edd}=1.5\times 10^{38}\left(\mbh/\sunm\right){\rm \ergs}$ is the Eddington 
luminosity, $L_{\rm bol}=\eta \dot{M}_{\bullet}c^2$ is the bolometric luminosity 
(see Table~\ref{tab:sed}), and $\eta_{0.1}=\eta/0.1$ is the radiative efficiency of the accretion 
disk. Using a BH mass of $\mbh=8.71\times 10^7\,\sunm$ and a mean bolometric luminosity 
of $L_{\rm bol}=10^{44.33} \, {\rm erg~s^{-1}}$ (Table~\ref{tab:sed}), 
we obtain $\mathdotM=0.21$ for $\eta=0.1$. The BH in NGC 5548 has an accretion 
rate in the regime of the standard accretion disk model (\citealt{Shakura1973}). 
Note that the above accretion rate is inaccurate if NGC 5548 hosts a binary supermassive BH, 
as recently suggested by  \cite{Li2016} based on the detection of periodic 
variations in luminosity and velocity.

\subsection{Potential Explanation: Radiation Pressure?}
\begin{figure}[t!]
\begin{center}
\includegraphics[angle=0,width=0.8\textwidth]{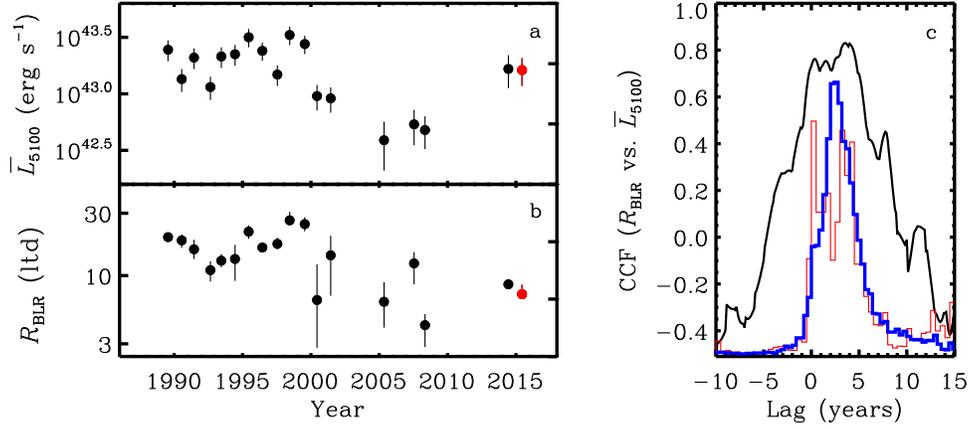}
\end{center}
\caption{\footnotesize
The secular variations of epoch-averaged ({\it a}) optical luminosity, $\bar{L}_{5100}$, and  ({\it b}) BLR size,
$R_{\rm BLR}$. The red point is from the present campaign.
Panel ({\it c}) shows the CCF of $\rblr$ with respect to $\bar{L}_{5100}$
and the distributions of Monte Carlo simulations of the centroid (blue) and peak (red) lags. 
}
\label{fig:obs_time}
\end{figure}

Figure~\ref{fig:obs_time} shows the variations of the mean $\bar{L}_{5100}$ and 
$\rblr$ over all the 18 campaigns since 1989. 
The variation amplitude of $\bar{L}_{5100}$ exceeds an order of magnitude. 
The lowest $L_{5100}\sim 10^{42.2}\,{\rm\ergs}$ occurred in 2005$-$2009 
(\citealt{Bentz2007}; \citealt{Bentz2009}; \citealt{Denney2010}) and 
the highest $L_{5100}\sim 10^{43.5}\,{\rm \ergs}$ in 1998--1999 
(\citealt{Peterson1999}; \citealt{Peterson2002}). Meanwhile, $\rblr$ also exhibits large variations, from $\sim 5$ up to $\sim 30$ light-days. 
Simple inspection of Figure~\ref{fig:obs_time} reveals that the changes in $\rblr$ follow the variations of $\bar{L}_{5100}$, but plausibly with a time delay.  Using the same procedure for computing H$\beta$ lags, we determine the lag of $\rblr$ with respect to $\bar{L}_{5100}$: $\tau_{_{R-\bar{L}}}=2.35_{-1.25}^{+3.47}$\,yrs. From analysis of the CCCD and CCPD, the probability of $\tau_{_{R-\bar{L}}}\le 0$\,yrs $p=0.06$ and 0.098, respectively, suggesting that there may be a potential delay between $\rblr$ and $\bar{L}_{5100}$. Clearly, 
we need more RM monitoring campaigns to verify this intriguing, but tentative result. 

There are two timescales for the ionized clouds in the BLR. The recombination timescale is 
$t_{\rm rec}=(n_e\alpha_{\rm B})^{-1}=6.0\,n_{10}^{-1}$\,min, where $n_{10}=n_e/10^{10}\,{\rm cm^{-3}}$ 
is the electron density of the clouds and $\alpha_{\rm B}$ is the case B recombination coefficient 
\citep{Osterbrock1989}. The fast response of the ionization front exactly follows the ionizing luminosity, as shown by Equation (6). However, radiation pressure, if effective, drives the clouds to outwards, changes their orbits, and leads to variations of their spatial distribution.  The observed size of the BLR is actually an emissivity-averaged value over the whole BLR, namely $\rblr=\int R\epsilon dR/\int \epsilon dR$, where $\epsilon$ is the emissivity of the clouds, determined by their spatial distribution (or, equivalently, their number density). Radiation pressure 
induces changes in $\epsilon$, and thereby $\rblr$. Considering that the H$\beta$ line
width represents the bulk motion of the clouds, the dynamical timescale of the BLR 
is given by (\citealt{Peterson1993}) 
\begin{equation}
 t_{_{\rm BLR}}=\frac {c\tau_{_{\rm H\beta}}} {V_{\rm FWHM}}=3.36\,\tau_{_{\rm 20}}V_{\rm 5000}^{-1}~{\rm yrs},
\label{eq:mir}
\end{equation}
where $V_{\rm 5000}=V_{\rm FWHM}/5000\,{\rm km~s^{-1}}$ and $\tau_{_{\rm 20}}=\tau_{_{\rm H\beta}}/20\,{\rm days}$. The quantity $t_{_{\rm BLR}}$ represents the typical timescale with which $\rblr$ varies in response to a change in the BLR dynamics.  For NGC~5548, the average H$\beta$ lag of $\tau_{_{\rm H\beta}}=15$ days and line width of $V_{\rm FWHM}$ = 6000 km s$^{-1}$ leads to a dynamical timescale of $t_{_{\rm BLR}}\approx 2.10$\,yrs. 

Interestingly, $t_{_{\rm BLR}}\approx \tau_{_{R-\bar{L}}}$, 
indicating that the BLR could be jointly controlled by radiation pressure and BH gravity.  
A mass inflow to the center can give rise to variations of the BLR and the accretion disk, 
but changes in the BLR structure (size) should precede changes in the disk luminosity. 
The delayed response of the BLR size to continuum luminosity may rule out this possibility, 
plausibly implicating the potential role of radiation pressure.

\section{Conclusions}
We present results of a new RM campaign on NGC~5548 based on high-quality optical spectra taken in 2015. We measure a centroid time lag for the broad $\rm H\beta$ line of $\tau_{_{\rm H\beta}} = 7.20^{+1.33}_{-0.35}$ days in the rest frame. 
Adopting a virial factor of $\fblr=6.3\pm1.5$ and 
an H$\beta$ line dispersion of $\sigma_{\rm line} = 3124\pm302$ \kms, we measured a BH mass of 
$\mbh=8.71^{+3.21}_{-2.61} \times 10^7M_{\odot}$.  We obtain the following results: 
\begin{itemize}
\item 
The velocity-resolved delay map of the broad H$\beta$ line shows a symmetric structure, consistent with the previous results of \cite{Denney2009b}. 

\item  
The relation between H$\beta$ line width and H$\beta$ time lag is consistent with virial motions.  
The virial product varies weakly with luminosity but is largely constant. 

\item
The BLR size of NGC 5548 follows $R_{\rm H\beta} \propto L_{\rm 5100}^{0.86}$,
steeper than the slope of $\sim 0.5$ for the global $R_{\rm H\beta}-L_{\rm 5100}$ relation for all RM AGNs.

\item 
Examining the variation patterns of $\rblr$ and $\bar{L}_{5100}$, 
we find tentative evidence that $\rblr$ follows $\bar{L}_{5100}$ 
with a delay of $2.35^{+3.47}_{-1.25}$\,yrs. 
This is consistent with the dynamical timescale of the BLR, 
implying that the long-term variations of the BLR may be driven by radiation pressure.

\end{itemize}

\acknowledgements{We are grateful to the referee for a helpful report.
We acknowledge the support of the staff of the Lijiang 2.4m telescope. 
Funding for the telescope has been provided by CAS and the People's Government of Yunnan 
Province. This research is supported by the Strategic Priority Research Program - The 
Emergence of Cosmological Structures of the Chinese Academy of Sciences, Grant No. XDB09000000, 
by NSFC grants NSFC-11173023, -11133006, -11373024, -11273007
-11233003, -11573026, -11503026 and -11473002, and a NSFC-CAS joint key grant U1431228.}

\clearpage

\appendix

\section{Mean and RMS Spectra}
\begin{figure}[t!]
\begin{center}
\includegraphics[angle=0,width=0.45\textwidth]{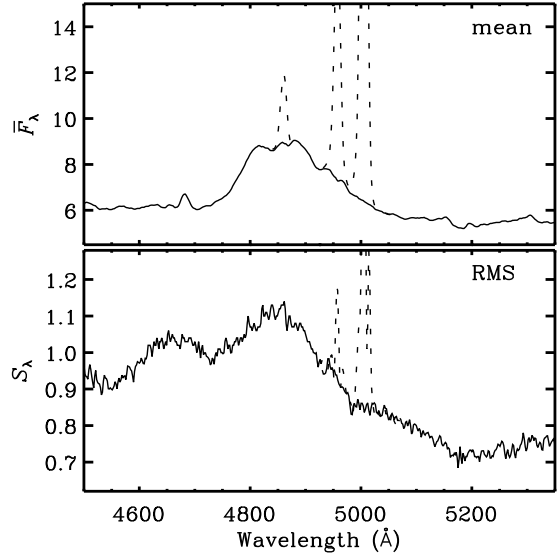}
\end{center}
\caption{\footnotesize
The mean ($\bar{F}_{\lambda}$) and RMS ($S_{\lambda}$) spectrum in units of 
$10^{-15}~{\rm erg~s^{-1}~cm^{-2}~{\textrm{\AA}}^{-1}}$ (solid line). 
The dash line represents the components of narrow emission lines.
}
\label{fig:mean_rms}
\end{figure}

The standard definitions of mean and RMS spectra are given by (\citealt{Peterson2004})
\begin{equation}
\bar{F}_{\lambda}=\frac{1}{N}\sum_{i=1}^NF_i(\lambda) 
\end{equation}
and
\begin{equation}
S_{\lambda}=\left\{\frac{1}{N}\sum_{i=1}^N\left[F_i(\lambda)-\overline{F}(\lambda)\right]^2\right\}^{1/2},
\end{equation}
where $F_i(\lambda)$ is the $i-$th spectrum and $N$ is the total number of spectra 
obtained during the campaign. We calculated the mean and RMS spectra of NGC 5548 
from all the spectra with absolute flux calibration and show them in Figure~\ref{fig:mean_rms}.
The line dispersion is calculated as
\begin{equation}
\sigma_{\rm line}^2(\lambda)=\langle\lambda^2\rangle-\lambda_0^2,
\end{equation}
where $\lambda_0=\int\lambda P(\lambda)d\lambda/\int P(\lambda) d\lambda$, 
$\langle\lambda^2\rangle=\int\lambda^2 P(\lambda)d\lambda/\int P(\lambda) d\lambda$, 
and $P(\lambda)$ is the line profile.

\clearpage

%
\begin{deluxetable}{lcccc}
\tablecolumns{5}
\tabletypesize{\scriptsize}
\tablewidth{0pt}
\tablecaption{Continuum and $\rm H\beta$ fluxes for NGC~5548}
\tablehead{
\colhead{JD}                           &
\colhead{$F_{\rm 5100}$}               &
\colhead{$F_{\rm H\beta}$}             &
\colhead{JD}                           &
\colhead{$V-$band}                     \\
\colhead{$-$2450000}                              &
\colhead{}  &
\colhead{}            &
\colhead{$-$2450000}                              &
\colhead{(mag)}                                  
}
\startdata
   7030.50 &$ 5.46 \pm 0.29 $&$ 8.00\pm 0.24 $&  7030.50 &$ 1.38 \pm 0.05$\\ 
   7037.50 &$ 4.96 \pm 0.29 $&$ 7.96\pm 0.24 $&  7037.50 &$ 1.45 \pm 0.01$\\
   7043.50 &$ 5.67 \pm 0.33 $&$ 7.89\pm 0.27 $&  7044.50 &$ 1.45 \pm 0.01$\\
   7047.50 &$ 4.65 \pm 0.32 $&$ 7.78\pm 0.26 $&  7046.50 &$ 1.44 \pm 0.01$\\
   7055.50 &$ 4.17 \pm 0.28 $&$ 7.84\pm 0.24 $&  7047.50 &$ 1.46 \pm 0.01$\\
\enddata
\tablecomments{\footnotesize 
$F_{\rm 5100}$ is the flux density at 
 5100 \AA\, in units of
$10^{-15}{\rm erg~s^{-1}~ cm^{-2}~\AA^{-1}}$ 
and $F_{\rm H\beta}$ is the H$\beta$ flux in units of 
$10^{-13}{\rm erg~s^{-1}~ cm^{-2}}$.
(This table is available in its entirety in a machine-readable form in 
the online journal. A portion is shown here for guidance regarding its form and content.)}
\label{tab:lc_dat}
\end{deluxetable}

\begin{deluxetable}{lcccccc}
\tablecolumns{7}
\tabletypesize{\scriptsize}
\tablewidth{0pt}
\tablecaption{Statistics of Light Curves for NGC~5548 in 2015}
\tablehead{
\colhead{Time Series}                    &
\colhead{$N$}                              &
\colhead{$\langle T\rangle$}             &
\colhead{$T_{\rm median}$}               &
\colhead{Mean Flux\tablenotemark{a}}     & 
\colhead{$F_{\rm var}$}                  &
\colhead{$R_{\rm max}$}                 \\
\colhead{}                               &
\colhead{}                               &
\colhead{(days)}                           &
\colhead{(days)}                           &
\colhead{} &
\colhead{}                                    &
\colhead{} \\
\colhead{(1)} &
\colhead{(2)} &
\colhead{(3)} &
\colhead{(4)} &
\colhead{(5)} &
\colhead{(6)} &
\colhead{(7)} 
}
\startdata
%
%
5100 \AA       & 62 & 3.4 &  3.0 &$4.34 \pm 0.30$ &$ 0.23 \pm 0.02 $&2.31 \\
$\rm H\beta$   & 62 & 3.4 &  3.0 &$6.95 \pm 0.25$ &$ 0.10 \pm 0.01 $&1.52 
\enddata
\tablecomments{\footnotesize
Col. (1) is time series. 
Col. (2) is the number of data points. 
Cols. (3) and (4) are the mean and median sampling intervals, respectively. 
Col. (5) is the mean flux and standard deviation. 
Cols. (6) and (7) are $F_{\rm var}$ and $R_{\rm max}$ defined in Section  3.1. \\
$^a$  The units of $F_{5100}$ and $F_{\rm H\beta}$ are the same as in Table~\ref{tab:lc_dat}.}
\label{tab:lcst}
\end{deluxetable}

%
\begin{deluxetable}{lc}
\tablecolumns{2}
\tabletypesize{\scriptsize}
\tablewidth{0pt}
\tablecaption{RM Measurements}
\tablehead{
\colhead{Parameter}  &
\colhead{Value}      
}
\startdata
$\tau_{\rm cent}$ ($\rm H\beta$ vs. $F_{5100}$)   & $7.20^{+1.33}_{-0.35}$ days\\
$\tau_{\rm peak}$ ($\rm H\beta$ vs. $F_{5100}$)   & $7.18^{+1.38}_{-0.70}$ days\\
FWHM (RMS)                                        & $ 9450\pm 290 $ km s$^{-1}$\\
$\sigma_{\rm line} (\rm RMS)$                     & $ 3124\pm 302 $ km s$^{-1}$\\
FWHM (mean)                                       & $ 9912\pm 362 $ km s$^{-1}$\\
$\sigma_{\rm line} (\rm mean)$                    & $ 3350\pm 272 $ km s$^{-1}$\\
$\log(\bar{L}_{\rm 5100}/\ergs$)                  & $43.21 \pm 0.12$ \\
$\log(\bar{L}_{\rm H\beta}/\ergs$)                & $41.70 \pm 0.05$ \\
\enddata
\tablecomments{ $\tau_{\rm cent}$ and $\tau_{\rm peak}$ are given in the rest-frame.}
\label{tab:rm_p}
\end{deluxetable}

\begin{deluxetable}{lccccccccccccc}
\rotate
  \tablecolumns{12}
  \tabletypesize{\scriptsize}
  \setlength{\tabcolsep}{3pt}
  \tablewidth{0pt}
  \tablecaption{All the RM measurements of NGC 5548 }
  \tablehead{
      \colhead{}&
  \colhead{}&
  \colhead{}&
  \colhead{}&
  \colhead{}&
  \colhead{}&
  \colhead{}&
  \multicolumn{2}{c}{Mean spectra}&
  \colhead{}&  
  \multicolumn{2}{c}{RMS spectra}&
  \colhead{}&
  \colhead{} \\ \cline{11-12} \cline{8-9} 
      \colhead{Data Set}&
  \colhead{Observation}                      &
  \colhead{$T$}                              &
  \colhead{$F_{\rm var}$}                    &
  \colhead{log $\bar L_{5100}$}              &
  \colhead{log $\bar L_{\rm H\beta}$}        &
  \colhead{H$\beta$ lags}                    & 
  \colhead{FWHM}                         &
  \colhead{$\sigma_{\rm line}$}          &  
    \colhead{}                           &
  \colhead{FWHM}                         &
  \colhead{$\sigma_{\rm line}$}          &
  \colhead{log $\mbh$}                   &
  \colhead{Ref.}                         \\ 
      \colhead{}&
  \colhead{Epoch}                             &
  \colhead{(Year)}                             &
  \colhead{}                                 &  
  \colhead{(erg~s$^{-1}$)}                   &
  \colhead{(erg~s$^{-1}$)}                   &
  \colhead{(days)}                                         &
  \colhead{(km s$^{-1}$)}                                  &
  \colhead{(km s$^{-1}$)}                                  & 
  \colhead{}                                               &  
  \colhead{(km s$^{-1}$)}                                  &
  \colhead{(km s$^{-1}$)}                                  &
  \colhead{($\sunm$)}                                      &
  \colhead{}     \\
      \colhead{(1)}&
  \colhead{(2)}        &
  \colhead{(3)}        &
  \colhead{(4)}        &
  \colhead{(5)}        &
  \colhead{(6)}        &
  \colhead{(7)}        &
  \colhead{(8)}        &
  \colhead{(9)}        & 
  \colhead{}           &
  \colhead{(10)}        &
  \colhead{(11)}       &
  \colhead{(12)}       &
  \colhead{(13)}       
} 
\startdata
Year 1  & 1988 Dec$-$1989 Oct  &1989.55& 0.188 &$  43.39\pm0.09 $&$ 41.76\pm0.04 $&$ 19.70 _{-1.50}^{+1.50} $&$ 4674  \pm  63 $&$1934 \pm 5   $&&$  4044  \pm 199  $&$ 1687 \pm  56 $&$ 7.84_{-0.11}^{+0.11}$& 1,2,3\\ 
Year 2  & 1989 Dec$-$1990 Oct  &1990.55& 0.272 &$  43.13\pm0.10 $&$ 41.61\pm0.05 $&$ 18.60 _{-2.30}^{+2.10} $&$ 5418  \pm 107 $&$2223 \pm 20  $&&$  4664  \pm 324  $&$ 1882 \pm  83 $&$ 7.91_{-0.12}^{+0.12}$& 1,2,3\\
Year 3  & 1990 Nov$-$1991 Oct  &1991.45& 0.154 &$  43.32\pm0.09 $&$ 41.69\pm0.06 $&$ 15.90 _{-2.50}^{+2.90} $&$ 5236  \pm  87 $&$2205 \pm 16  $&&$  5776  \pm 237  $&$ 2075 \pm  81 $&$ 7.93_{-0.13}^{+0.13}$& 1,2,3\\ 
Year 4  & 1992 Jan$-$1992 Oct  &1992.65& 0.386 &$  43.06\pm0.10 $&$ 41.48\pm0.05 $&$ 11.00 _{-2.00}^{+1.90} $&$ 5986  \pm  95 $&$3109 \pm 53  $&&$  5691  \pm 164  $&$ 2264 \pm  88 $&$ 7.84_{-0.13}^{+0.13}$& 1,2,3\\ 
Year 5  & 1992 Nov$-$1993 Sep  &1993.45& 0.148 &$  43.33\pm0.09 $&$ 41.73\pm0.06 $&$ 13.00 _{-1.40}^{+1.60} $&$ 5930  \pm  42 $&$2486 \pm 13  $&&$         -       $&$ 1909 \pm 129 $&$ 7.77_{-0.13}^{+0.13}$& 1,2,3\\ 
Year 6  & 1993 Nov$-$1994 Oct  &1994.45& 0.173 &$  43.35\pm0.09 $&$ 41.69\pm0.04 $&$ 13.40 _{-4.30}^{+3.80} $&$ 7378  \pm  39 $&$2877 \pm 17  $&&$  7202  \pm 392  $&$ 2895 \pm 114 $&$ 8.14_{-0.18}^{+0.16}$& 1,2,3\\
Year 7  & 1994 Nov$-$1995 Oct  &1995.45& 0.117 &$  43.50\pm0.08 $&$ 41.80\pm0.04 $&$ 21.70 _{-2.60}^{+2.60} $&$ 6946  \pm  79 $&$2432 \pm 13  $&&$  6142  \pm 289  $&$ 2247 \pm 134 $&$ 8.13_{-0.13}^{+0.13}$& 1,2,3\\
Year 8  & 1995 Nov$-$1996 Oct  &1996.45& 0.244 &$  43.38\pm0.08 $&$ 41.73\pm0.04 $&$ 16.40 _{-1.10}^{+1.20} $&$ 6623  \pm  93 $&$2276 \pm 15  $&&$  5706  \pm 357  $&$ 2026 \pm  68 $&$ 7.92_{-0.11}^{+0.11}$& 1,2,3\\ 
Year 9  & 1996 Dec$-$1997 Oct  &1997.55& 0.209 &$  43.17\pm0.09 $&$ 41.67\pm0.09 $&$ 17.50 _{-1.60}^{+2.00} $&$ 6298  \pm  65 $&$2178 \pm 12  $&&$  5541  \pm 354  $&$ 1923 \pm  62 $&$ 7.90_{-0.11}^{+0.12}$& 1,2,3\\ 
Year 10 & 1997 Nov$-$1998 Sep  &1998.45& 0.146 &$  43.52\pm0.08 $&$ 41.86\pm0.03 $&$ 26.50 _{-2.20}^{+4.30} $&$ 6177  \pm  36 $&$2035 \pm 11  $&&$  4596  \pm 505  $&$ 1732 \pm  76 $&$ 7.99_{-0.12}^{+0.13}$& 1,2,3\\ 
Year 11 & 1998 Nov$-$1999 Oct  &1999.55& 0.229 &$  43.44\pm0.08 $&$ 41.76\pm0.06 $&$ 24.80 _{-3.00}^{+3.20} $&$ 6247  \pm  57 $&$2021 \pm 18  $&&$  6377  \pm 147  $&$ 1980 \pm  30 $&$ 8.08_{-0.12}^{+0.12}$& 1,2,3\\ 
Year 12 & 1999 Dec$-$2000 Sep  &2000.45& 0.424 &$  42.98\pm0.11 $&$ 41.57\pm0.04 $&$ 6.50  _{-3.70}^{+5.70} $&$ 6240  \pm  77 $&$2010 \pm 30  $&&$  5957  \pm 224  $&$ 1969 \pm  48 $&$ 7.49_{-0.27}^{+0.40}$& 1,2,3$^{a}$\\ 
Year 13 & 2000 Nov$-$2001 Dec  &2001.45& 0.293 &$  42.96\pm0.11 $&$ 41.46\pm0.05 $&$ 14.30 _{-7.30}^{+5.90} $&$ 6478  \pm 108 $&$3111 \pm 131 $&&$  6247  \pm 343  $&$ 2173 \pm  89 $&$ 7.92_{-0.17}^{+0.21}$& 1,2,3\\ 
Year 17 & 2005 Mar$-$2005 Apr  &2005.35& 0.187 &$  42.59\pm0.20 $&$ 41.07\pm0.09 $&$ 6.30  _{-2.30}^{+2.60} $&$ 6396  \pm 167 $&$3210 \pm 642 $&&$         -       $&$ 2939 \pm 768 $&$ 7.83_{-0.22}^{+0.23}$& 1,4\\ 
Year 19 & 2007 Mar$-$2007 Jul  &2007.55& 0.157 &$  42.73\pm0.15 $&$ 41.19\pm0.10 $&$ 12.40 _{-3.85}^{+2.74} $&$ 11481 \pm 574 $&$      -      $&&$  4849  \pm 112  $&$ 1822 \pm  35 $&$ 7.70_{-0.17}^{+0.14}$& 1,5\\  
Year 20 & 2008 Feb$-$2008 Jun  &2008.35& 0.227 &$  42.68\pm0.14 $&$ 41.21\pm0.06 $&$ 4.17  _{-1.33}^{+0.90} $&$ 12771 \pm  71 $&$4266 \pm  65 $&&$  11177 \pm 2266 $&$ 4270 \pm 292 $&$ 7.97_{-0.18}^{+0.15}$& 1,6$^{b}$\\  
Year 25 & 2013 Dec$-$2014 Aug  &2014.45&  $-$  &$  43.22\pm0.14 $&$       -      $&$ 8.57  _{-0.67}^{+0.67} $&$        -      $&$      -      $&&$         -       $&$        -     $&$          -          $& 7\\  
Year 26 & 2015 Jan$-$2015 Jul  &2015.45& 0.233 &$  43.21\pm0.12 $&$ 41.70\pm0.05 $&$ 7.20  _{-0.35}^{+1.33} $&$ 9912  \pm 362 $&$3350 \pm 272 $&&$  9450  \pm 290  $&$ 3124 \pm 302 $&$ 7.94_{-0.13}^{+0.16}$& 8
\enddata
\tablecomments{\footnotesize 
$\mbh$ is calculated from the line dispersion of the RMS spectrum with a virial factor of $f_{\rm BLR}=6.3\pm1.5$ (\citealt{Ho2014}). These campaigns yield a mean BH mass of  
$\langle \mbh\rangle=(8.39\pm 0.72)\times10^{7}\sunm$. \\
$^{a}$  ``Year 12"  is compiled from the 12th RM observation in the AGN Watch project (\citealt{Peterson2002}). 
The cross-correlation analysis is very ambiguous 
(see Figure 2 of \citealt{Peterson2002}), and we excluded this data set in our analysis. \\
$^{b}$ \citet{Pancoast2014} modelled the data and provided 
the model-dependent H$\beta$ lag of $3.22_{-0.54}^{+0.66}$ days, which is consistent (within the uncertainties) 
with $4.17_{-1.33}^{+0.90}$ days determined by the cross correlation analysis in \cite{Bentz2010}. 
We use the H$\beta$ lag of \citet{Bentz2010} for consistency.   \\
{\bf References.} 
(1) \citet{Eser2015}, 
(2) \citet{Collin2006}, 
(3) \citet{Peterson2004}, 
(4) \citet{Bentz2007}, 
(5) \citet{Denney2010},
(6) \citet{Bentz2010},
(7) \citet{Fausnaugh2015arXiv}, 
(8) This work.
}
\label{tab:tot}
\end{deluxetable}

\begin{deluxetable}{llccc}
\tablecolumns{2}
\tabletypesize{\scriptsize}
\tablewidth{0pt}
\tablecaption{Simultaneous Observations of the Spectral Energy Distribution for NGC 5548}
\tablehead{
\colhead{References}                           &
\colhead{Observations}                         &
\colhead{$\log \left(L_{\rm bol}/\ergs\right)$}&
\colhead{$\epsilon_{_{\rm Edd}}$}              &
\colhead{Notes}
}
\startdata
V09a &XMM(OM, EPIC-pn) in Dec. 2000            & $44.35\pm0.04$ & $0.018\pm0.002$ & simultaneous\\
V09b &Swift(UVOT, XRT) in Jul. 2007            & $44.15\pm0.04$ & $0.011\pm0.002$ & simultaneous \\
V10  &Swift(BAT) in Jul. 2007, IRAS in 1983    & $44.50\pm0.04$ & $0.025\pm0.003$ &not simultaneous
\enddata
\tablecomments{\footnotesize 
{\bf References}: V09a: \cite{Vasudevan2009a}; V09b: \cite{Vasudevan2009b}; V10: \cite{Vasudevan2010}.\\
There are many observations of NGC 5548 (see \citealt{Chiang2003} for a brief summary of data).
Here we only list the recent observations. ``Simultaneous" means that the UV, optical, and X-ray data
are simultaneously observed.
}
\label{tab:sed}
\end{deluxetable}

\clearpage

%

%

%
\clearpage

%

\begin{thebibliography}{99}
\bibitem[Ade et al.(2014)]{Ade2014}
   Ade, P.~A.~R., Arnaud, M., et al. (Planck Collaboration) \ 2014, \aap, 571, A31 
\bibitem[Alexander(1997)]{Alexander1997} 
   Alexander, T.\ 1997, Astronomical Time Series, eds. D. Maoz, A. Sternberg \& E.M. Leibowitz, Kluwer, Dordrecht, p. 163
\bibitem[Bahcall et al.(1972)]{Bahcall1972} 
   Bahcall, J.~N., Kozlovsky, B.-Z., \& Salpeter, E.~E.\ 1972, \apj, 171, 467 
\bibitem[Bentz et al.(2013)]{Bentz2013} 
   Bentz, M.~C., Denney, K.~D., Grier, C.~J., et al.\ 2013, \apj, 767, 149 
\bibitem[Bentz et al.(2010)]{Bentz2010} 
   Bentz, M.~C., Horne, K., Barth, A.~J., et al.\ 2010, \apjl, 720, L46   
\bibitem[Bentz et al.(2009)]{Bentz2009} 
   Bentz, M.~C., Walsh, J.~L., Barth, A.~J., et al.\ 2009, \apj, 705, 199 
\bibitem[Bentz et al.(2008)]{Bentz2008} 
   Bentz, M.~C., Walsh, J.~L., Barth, A.~J., et al.\ 2008, \apjl, 689, L21
\bibitem[Bentz et al.(2007)]{Bentz2007}
   Bentz, M.~C., Denney, K.~D., Cackett, E.~M., et al.\ 2007, \apj, 662, 205 
\bibitem[Blandford \& McKee(1982)]{Blandford1982} 
   Blandford, R.~D., \& McKee, C.~F.\ 1982, \apj, 255, 419 
\bibitem[Boroson \& Green(1992)]{Boroson1992} 
   Boroson, T.~A., \& Green, R.~F.\ 1992, \apjs, 80, 109 
\bibitem[Bruzual \& Charlot(2003)]{Bruzual2003} 
   Bruzual, G., \& Charlot, S.\ 2003, \mnras, 344, 1000 
\bibitem[Chiang \& Blaes(2003)]{Chiang2003}
   Chiang, J. \& Blaes, O. 2003, \apj, 586, 97
\bibitem[Collin et al.(2006)]{Collin2006} 
   Collin, S., Kawaguchi, T., Peterson, B.~M., \& Vestergaard, M.\ 2006, \aap, 456, 75
\bibitem[Denney et al.(2010)]{Denney2010} 
   Denney, K.~D., Peterson, B.~M., Pogge, R.~W., et al.\ 2010, \apj, 721, 715  
\bibitem[Denney et al.(2009{\natexlab{a}})]{Denney2009a} 
   Denney, K.~D., Watson, L.~C., Peterson, B.~M., et al.\ 2009{\natexlab{a}}, \apj, 702, 1353
\bibitem[Denney et al.(2009{\natexlab{b}})]{Denney2009b} 
   Denney, K.~D., Peterson, B.~M., Pogge, R.~W., et al.\ 2009{\natexlab{b}}, \apjl, 704, L80
\bibitem[De Rosa et al.(2015)]{DeRosa2015}
   De Rosa, G., Peterson, B.~M., Ely, J., et al.\ 2015, \apj, 806, 128
\bibitem[Du et al. (2016a)]{Du2016a}
   Du, P., Lu, K.-X., Hu, C., et al.\ 2016, \apj, submitted 
\bibitem[Du et al.(2016b)]{Du2016b} 
   Du, P., Lu, K.-X., Hu, C., et al.\ 2016, \apj, 820, 27 
\bibitem[Du et al.(2015)]{Du2015} 
   Du, P., Hu, C., Lu, K.-X., et al. 2015, \apj, 806, 22
\bibitem[Du et al.(2014)]{Du2014} 
   Du, P., Hu, C., Lu, K.-X., et al. 2014, \apj, 782, 45
\bibitem[Edelson et al.(2015)]{Edelson2015}
   Edelson, R., Gelbord, J. M., Horne, K., et al. 2015, \apj, 806, 129
\bibitem[Edelson et al.(2002)]{Edelson2002} 
   Edelson, R., Turner, T.~J., Pounds, K., et al.\ 2002, \apj, 568, 610 
\bibitem[Edelson \& Krolik(1988)]{Edelson1988}
   Edelson, R.~A., \& Krolik, J.~H.\ 1988, \apj, 333, 646 
\bibitem[Fausnaugh et al.(2015)]{Fausnaugh2015arXiv} 
   Fausnaugh, M.~M., Denney, K.~D., Barth, A.~J., et al.\ 2015, arXiv:1510.05648 
\bibitem[Filippenko(1982)]{Filippenko1982}
   Filippenko, A. 1982, \pasp, 94, 715
\bibitem[Gaskell \& Peterson(1987)]{Gaskell1987} 
   Gaskell, C.~M., \& Peterson, B.~M.\ 1987, \apjs, 65, 1
\bibitem[Gaskell \& Sparke(1986)]{Gaskell1986} 
   Gaskell, C.~M., \& Sparke, L.~S.\ 1986, \apj, 305, 175
\bibitem[Grier et al.(2013)]{Grier2013} 
   Grier, C.~J., Peterson, B.~M., Horne, K., et al.\ 2013, \apj, 764, 47
\bibitem[Ho \& Kim(2014)]{Ho2014} 
   Ho, L.~C., \& Kim, M.\ 2014, \apj, 789, 17 
\bibitem[Hu et al.(2015)]{Hu2015} 
   Hu, C., Du, P., Lu, K.-X., et al. 2015, \apj, 804, 138  
\bibitem[Kaspi et al.(2000)]{Kaspi2000} 
   Kaspi, S., Smith, P.~S., Netzer, H., et al.\ 2000, \apj, 533, 631
\bibitem[Kaspi et al.(2005)]{Kaspi2005} 
   Kaspi, S., Maoz, D., Netzer, H., et al.  2005, \apj, 629, 61
\bibitem[Kilerci Eser et al.(2015)]{Eser2015} 
   Kilerci Eser, E., Vestergaard, M., Peterson, B.~M., et al. 2015, \apj, 801, 8 
\bibitem[Kollatschny \& Zetzl(2013)]{Kollatschny2013}
   Kollatschny, W. \& Zetzl, M. 2013, \aap, 549, A100 
\bibitem[Kormendy \& Ho(2013)]{Kormendy2013}
   Kormendy, J., \& Ho, L.~C.\ 2013, \araa, 51, 511 
\bibitem[Li et al.(2013)]{Li2013}
   Li, Y.-R., Wang, J.-M., Ho, L.~C., Du, P., \& Bai, J.-M.\ 2013, \apj, 779, 110 
\bibitem[Li et al.(2016)]{Li2016} 
   Li, Y.-R., Wang, J.-M., Ho, L.~C., et al.\ 2016, \apj\, in press (arXiv:1602.05005)
\bibitem[Maoz et al.(1990)]{Maoz1990} 
   Maoz, D., Netzer, H., Leibowitz, E., et al.\ 1990, \apj, 351, 75 
\bibitem[Onken et al.(2004)]{Onken2004} 
   Onken, C.~A., Ferrarese, L., Merritt, D., et al.\ 2004, \apj, 615, 645
\bibitem[Osterbrock (1989)]{Osterbrock1989}
   Osterbrock, D. E. 1989, Astrophysics of Gaseous Nebulae and Active Galactic Nuclei (Mill Valley, University Science Books)
\bibitem[Pancoast et al.(2011)]{Pancoast2011}
   Pancoast, A., Brewer, B. J. \& Treu, T. 2011, \apj, 730, 139
\bibitem[Pancoast et al.(2014a)]{Pancoast2014a} 
   Pancoast, A., Brewer, B.~J., \& Treu, T.\ 2014a, \mnras, 445, 3055 
\bibitem[Pancoast et al.(2014b)]{Pancoast2014} 
   Pancoast, A., Brewer, B.~J., Treu, T., et al.\ 2014b, \mnras, 445, 3073
\bibitem[Peterson(1993)]{Peterson1993}
   Peterson, B. 1993, \pasp, 105, 247
\bibitem[Peterson et al.(2004)]{Peterson2004} 
   Peterson, B.~M., Ferrarese, L., Gilbert, K.~M., et al.\ 2004, \apj, 613, 682  
\bibitem[Peterson et al.(2002)]{Peterson2002} 
   Peterson, B.~M., Berlind, P., Bertram, R., et al.\ 2002, \apj, 581, 197 
\bibitem[Peterson et al.(1999)]{Peterson1999} 
   Peterson, B.~M., Barth, A.~J., Berlind, P., et al.\ 1999, \apj, 510, 659 
\bibitem[Peterson et al.(1998)]{Peterson1998} 
   Peterson, B.~M., Wanders, I., Horne, K., et al.\ 1998, \pasp, 110, 660 
\bibitem[Press et al.(1992)]{Press1992}
   Press, W. H., Teukolsky, S. A., Vetterling, W. T., \& Flannery, B. P. 1992, Numerical Recipes in FORTRAN (2nd ed.; Cambridge: Cambridge Univ. Press)
\bibitem[Rodr{\'{\i}}guez-Pascual et al.(1997)]{Rodriguez-Pascual1997} 
   Rodr{\'{\i}}guez-Pascual, P.~M., Alloin, D., Clavel, J., et al.\ 1997, \apjs, 110, 9 
\bibitem[Sergeev et al.(2007)]{Sergeev2007}
   Sergeev, S.~G., Doroshenko, V.~T., Dzyuba, S.~A., et al.\ 2007, \apj, 668, 708 
\bibitem[Shakura \& Sunyaev(1973)]{Shakura1973}
   Shakura, N. I. \& Sunyaev, R. A. 1973, \aap, 24, 337 
\bibitem[Vasudevan \& Fabian(2009a)]{Vasudevan2009a} 
   Vasudevan, R.~V., \& Fabian, A.~C.\ 2009a, \mnras, 392, 1124
\bibitem[Vasudevan et al.(2009b)]{Vasudevan2009b} 
   Vasudevan, R.~V., Mushotzky, R.~F., Winter, L.~M., \& Fabian, A.~C.\ 2009b, \mnras, 399, 1553
\bibitem[Vasudevan et al.(2010)]{Vasudevan2010}
   Vasudevan, R.~V., Fabian, A.~C., Gandhi, P., Winter, L.~M., \& Mushotzky, R.~F.\ 2010, \mnras, 402, 1081
\bibitem[Wanders \& Peterson(1996)]{Wanders1996}
   Wanders, I., \& Peterson, B.~M.\ 1996, \apj, 466, 174 
\bibitem[White \& Peterson(1994)]{White1994} 
   White, R.~J., \& Peterson, B.~M.\ 1994, \pasp, 106, 879
\bibitem[Woo et al.(2010)]{Woo2010} 
   Woo, J.-H., Treu, T., Barth, A.~J., et al.\ 2010, \apj, 716, 269 
\end{thebibliography}
\end{document}